\documentstyle[11pt,epsf,psfig, amsmath]{article}

\def\Year{\expandafter\eatPrefix\the\year}
\newcount\hours \newcount\minutes
\def\monthname{\ifcase\month\or
January\or February\or March\or April\or May\or June\or July\or
August\or September\or October\or November\or December\fi}
\def\shortmonthname{\ifcase\month\or
Jan\or Feb\or Mar\or Apr\or May\or Jun\or Jul\or
Aug\or Sep\or Oct\or Nov\or Dec\fi}

\def\TimeStamp{\hours\the\time\divide\hours by60%
\minutes -\the\time\divide\minutes by60\multiply\minutes by60%
\advance\minutes by\the\time%
${\rm \shortmonthname}\cdot   \if\day<10{}0\fi\the\day\cdot   \the\year
\qquad\the\hours:\if\minutes<10{}0\fi\the\minutes$}

\newskip\humongous \humongous=0pt plus 1000pt minus 100pt
\def\caja{\mathsurround=0pt}
\def\eqalign#1{\,\vcenter{\openup1\jot \caja
       \ialign{\strut \hfil$\displaystyle{##}$&$
        \displaystyle{{}##}$\hfil\crcr#1\crcr}}\,}
\newif\ifdtup

\newcounter{eqnumber}[section]

\def\eqn#1{eq.~(\ref{#1})}

\def\eqns#1#2{eqs.~(\ref{#1}) and~(\ref{#2})}

\def\tr{\mathop{\rm tr}\nolimits}

\newbox\charbox
\newbox\slabox
\def\s#1{{      
        \setbox\charbox=\hbox{$#1$}
        \setbox\slabox=\hbox{$/$}
        \dimen\charbox=\ht\slabox
        \advance\dimen\charbox by -\dp\slabox
        \advance\dimen\charbox by -\ht\charbox
        \advance\dimen\charbox by \dp\charbox
        \divide\dimen\charbox by 2
        \raise-\dimen\charbox\hbox to \wd\charbox{\hss/\hss}
        \llap{$#1$}
}}

\def\spa#1.#2{\left\langle#1\,#2\right\rangle}
\def\spb#1.#2{\left[#1\,#2\right]}
\def\lor#1.#2{\left(#1\,#2\right)}

\catcode`@=11  

\def\Slash#1{\slash\hskip -0.22 cm #1}

\def\eps{\epsilon}

\def\e{\epsilon}
\def\lr{\leftrightarrow}

\def\la{\langle}
\def\ra{\rangle}

\def\lsl{\not{\hbox{\kern-2.3pt $\ell$}}}
\def\Psl{\not{\hbox{\kern-2.3pt $P$}}}
\def\ksl{\not{\hbox{\kern-2.3pt $k$}}}
\def\twosl{\not{\hbox{\kern-2.3pt $2$}}}
\def\fivesl{\not{\hbox{\kern-2.3pt $5$}}}

\def\spa#1.#2{\left\langle#1\,#2\right\rangle}
\def\spb#1.#2{\left[#1\,#2\right]}
\def\lor#1.#2{\left(#1\,#2\right)}

\def\sand#1.#2.#3{%
  \left\langle\smash{#1}{\vphantom1}\right|{#2}%
  \left|\smash{#3}{\vphantom1}\right\rangle}
\def\sandp#1.#2.#3{%
  \left\langle\smash{#1}{\vphantom1}^{-}\right|{#2}%
  \left|\smash{#3}{\vphantom1}^{+}\right\rangle}
\def\sandpp#1.#2.#3{%
  \left\langle\smash{#1}{\vphantom1}^{+}\right|{#2}%
  \left|\smash{#3}{\vphantom1}^{+}\right\rangle}
\def\sandmm#1.#2.#3{%
  \left\langle\smash{#1}{\vphantom1}^{-}\right|{#2}%
  \left|\smash{#3}{\vphantom1}^{-}\right\rangle}
\def\sandpm#1.#2.#3{%
  \left\langle\smash{#1}{\vphantom1}^{+}\right|{#2}%
  \left|\smash{#3}{\vphantom1}^{-}\right\rangle}
\def\sandmp#1.#2.#3{%
  \left\langle\smash{#1}{\vphantom1}^{-}\right|{#2}%
  \left|\smash{#3}{\vphantom1}^{+}\right\rangle}

\def\Atree{A^{\rm tree}}

\def\dlips{dLIPS}

\def\Lz{\mathop{\hbox{\rm L}}\nolimits_0}
\def\Kz{\mathop{\hbox{\rm K}}\nolimits_0}

\def\tr{\mathop{\hbox{\rm tr}}\nolimits}

\def\dlips{d{\rm LIPS}}
\def\Split{\mathop{\rm Split}\nolimits}

\begin{document}

\begin{titlepage}

\begin{flushright}
hep-th/0410296
\\
SWAT--04--416 \\
SLAC--PUB--10828
\end{flushright}

\vskip 2.cm

\begin{center}
\begin{Large}
{\bf $N=1$ Supersymmetric One-loop Amplitudes \\
  and the Holomorphic Anomaly of Unitarity Cuts}

\vskip 2.cm

\end{Large}

\vskip 2.cm

{\large
Steven J. Bidder${}^{1,\dagger}$,  N.\ E.\ J.\ Bjerrum-Bohr${}^{1,\dagger}$, Lance\ J. Dixon$^{2,\sharp}$ 
and David C. Dunbar$^{1,\dagger}$
} 

\vskip 0.5cm

{\it ${}^1$ Department of Physics \\
University of Wales Swansea \\ 
 Swansea, SA2 8PP, UK }

{\it $^{2}$ Stanford Linear Accelerator Center\\
 Stanford University\\
Stanford, CA 94309, USA}

\vskip .3cm

\begin{abstract}
Recently, it has been shown that the holomorphic anomaly of unitarity cuts
can be used as a tool in determining the one-loop amplitudes in $N=4$
super Yang-Mills theory.  It is interesting to examine whether this method
can be applied to more general cases.  We present results for a non-MHV $N=1$
supersymmetric one-loop amplitude.  We show that the holomorphic anomaly
of each unitarity cut correctly reproduces the action on the amplitude's
imaginary part of the differential operators corresponding to
collinearity in twistor space.  We find that the use of the holomorphic
anomaly to evaluate the amplitude requires the solution of differential
rather than algebraic equations.
\end{abstract}

\end{center}

\vfill
\noindent\hrule width 3.6in\hfil\break
${}^{\dagger}$ Research supported by the PPARC.\hfil\break
${}^{\sharp}$ Research supported by the US Department of Energy under contract
DE-AC02-76SF00515. \hfil\break

\end{titlepage}


\section{Introduction}

It has been proposed recently that a ``weak-weak'' duality exists
between $N=4$ supersymmetric gauge theory and topological string
theory~\cite{Witten:2003nn}. This relationship 
becomes manifest by transforming amplitudes into twistor space where 
they are supported on simple curves.
A consequence of this picture is that tree amplitudes,
when expressed in terms of spinor variables 
$k_{a\dot a} =\lambda_a \tilde\lambda_{\dot a}$, 
are annihilated by various differential operators corresponding 
to the localization of points upon lines and planes in twistor space.

In particular the operator corresponding to collinearity of points $i,j,k$
in twistor space,
\begin{equation}
[F_{ijk} , \eta ] =
 \spa{i}.j \left[{ \partial\over \partial\tilde\lambda_k},\eta\right]
+\spa{j}.k \left[{ \partial\over
\partial\tilde\lambda_i},\eta\right] +\spa{k}.i \left[{
\partial\over
\partial\tilde\lambda_j},\eta\right]
\end{equation}
annihilates  the
``maximally helicity violating'' (MHV) $n$-gluon amplitudes,
\begin{equation}
[F_{ijk} , \eta ] A_n^{\rm MHV\hbox{-}tree}(1,2,\cdots n) = 0.
\end{equation}
These MHV colour-ordered amplitudes, where exactly two of the gluons have
negative helicity, have a remarkably simple form, conjectured by Parke
and Taylor~\cite{ParkeTaylor} and proven by Berends and Giele~\cite{BerendsGiele},
\begin{equation}
A_n^{\rm MHV\hbox{-}tree}( 1^+,2^+,\ldots ,p^-,\ldots ,q^-,\ldots, n^+) 
=  i { \spa{p}.q^4  \over \spa1.2\spa2.3\spa3.4 \cdots \spa{n-1,}.n \spa{n}.1} \,.
\end{equation}
Using ``cut-constructibility'' and collinear limits, the one-loop MHV
amplitudes have also been constructed for $N=4$~\cite{BDDKa} and 
$N=1$ supersymmetric theories~\cite{BDDKb}. 
``Cut-constructible'' implies that the entire amplitude can be
reconstructed from a knowledge of its four-dimensional cuts
--- cuts evaluated with the intermediate states labelled by 
four-dimensional helicities~\cite{BDDKa,BDDKb}.

The MHV tree amplitudes appear to play a key role in gauge
theories and recently Cachazo, Svr\v{c}ek and Witten conjectured that
Yang-Mills amplitudes could be calculated using off-shell MHV
vertices~\cite{Cachazo:2004kj}.  Points localised on a single line
become attached to a MHV vertex.  There have already been multiple 
applications of the CSW construction to the calculation of tree 
amplitudes~\cite{Tree}.

Brandhuber {\it et al.}~\cite{Brandhuber:2004yw} emphasised the
usefulness of the twistor picture beyond tree level
by computing with MHV vertices and reproducing the one-loop $N=4$ 
MHV amplitudes.  The steps in their computation closely paralleled the
cut-based computation~\cite{BDDKa} (though the MHV vertices are
off-shell), stressing the key role of unitarity.  
Despite this success, application of the collinear and planar
operators to one-loop MHV amplitudes does not result in their
annihilation~\cite{Cachazo:2004zb}, even though naively it should.
More specifically, consider the operator $F_{i\; i+1\; i+2}$ 
acting upon the cut $C_{i,\ldots,j}$
--- the imaginary part, or $1/2$ the discontinuity,
in the channel $(k_i+k_{i+1}+\cdots + k_{j})^2 > 0$ --- 
of a one-loop MHV amplitude,
\begin{equation}
C_{i,\ldots,j}  \equiv  
{ i \over 2} \int \dlips\biggl[ A^{\rm MHV\hbox{-}tree}(\ell_1,i,i+1,\ldots,
j,\ell_2) \times A^{\rm MHV\hbox{-}tree}(-\ell_2,j+1,j+2,\ldots,i-1,-\ell_1)
\biggr] \,.
\end{equation}
Because $F_{i\; i+1\; i+2}$ annihilates both tree amplitudes, we might
naively expect it to annihilate the imaginary part of the amplitude ---
but it explicitly does not. (This computation has recently been extended to the
special case of $N=1$ MHV one-loop amplitudes~\cite{Brandhuber:new}.) 

This apparent paradox was resolved~\cite{Cachazo:2004by} by
observing that differential operators acting within the loop-momentum
integral yield $\delta$ functions. This ``holomorphic anomaly of the
unitarity cut'' produces a rational function as a result even though
the tree amplitudes within the cut are
localized on lines~\cite{Cachazo:2004by, Bena:2004xu,Cachazo:2004dr}.  
As a spin-off of this resolution, it
was observed that acting with $F_{ijk}$ upon both the cut and the
imaginary part of the amplitude, and demanding consistency, leads to
algebraic equations for the coefficients of the integral functions which
appear in the amplitude. These algebraic equations are helpful in
computing the entire amplitude~\cite{Cachazo:2004dr}, as
demonstrated in the calculation of one of the four
seven-point non-MHV $N=4$ amplitudes~\cite{Britto:2004nj}. The result
agrees with that of ref.~\cite{BeDeDiKo}, where this and the remaining 
three amplitudes were evaluated directly from the cuts.

In this letter we present the results for a six-gluon non-MHV one-loop
amplitude, where an $N=1$ chiral multiplet propagates in the loop.
This amplitude,
$A^{N=1\ {\rm chiral}}(1^-,2^-,3^-,4^+,5^+,6^+)$, was calculated by
conventional (cut-based) means.
We examine the action of the $F_{ijk}$ operator
upon the cuts of this amplitude, and demonstrate the consistency of
the holomorphic anomaly with the result of $F_{ijk}$ acting upon its
imaginary part.  We also investigate how the holomorphic anomaly could
be used to calculate $N=1$ one-loop amplitudes.  We find in general,
that the coefficients of the integral functions must now satisfy
differential rather than algebraic equations. We also examine how
coefficients of integral functions may be determined by these differential
equations, together with other constraints, {\it e.g.,} collinear limits, 
by considering the $n$-point amplitude, 
$A^{N=1\ {\rm chiral}}(1^-,2^-,3^-,4^+,\ldots,n^+)$.


\section{Organisation of $N=1$ Supersymmetric Amplitudes}
 
The computation of (unrenormalized) one-loop amplitudes may be simplified by
carefully considering the integral functions $I_i$ that may appear, and by 
realizing that the full amplitude is a linear combination of such
functions with rational (in the variables $\lambda_a^i$ and 
$\tilde\lambda_{\dot a}^j$) coefficients $c_i$,
\begin{equation}
A = \sum_{i}  c_i  I_i \,.
\label{generalform}
\end{equation}
For supersymmetric amplitudes the summation is over a restricted set of functions. 
In $N=4$ theories the functions that appear are only scalar box functions;
whereas for $N=1$ theories we
are limited to scalar boxes, scalar triangles and scalar bubbles.
In~\cite{BDDKa,BDDKb} it was demonstrated that both amplitudes
are  ``cut-constructible'', {\it i.e.,} the coefficients $c_i$ can be
entirely determined by knowledge of the four-dimensional cuts of the amplitude.

For $N=1$ super Yang-Mills with external gluons there are two
possible supermultiplets contributing to the loop amplitude ---
the vector and the chiral matter multiplets.  For simplicity we consider 
the leading-in-colour components of colour-ordered one-loop amplitudes. 
These can be decomposed into the contributions from single particle spins,
\begin{equation}
\eqalign{ A_{n}^{N=1\ {\rm vector}}\ \equiv\ A_{n}^{[1]}\
+A_{n}^{[1/2]} \,, \cr A_{n}^{N=1\ {\rm chiral}}\ \equiv\
A_{n}^{[1/2]}\ +A_{n}^{[0]} \,, \cr}
\end{equation}
where  $A_{n}^{[J]}$ is the one-loop amplitude with $n$ external 
gluons and particles of spin-$J$ circulating in the loop. 
(For spin-$0$ we mean a complex scalar.) For $N=4$ 
super Yang-Mills theory there is a single multiplet 
which is given by
\begin{equation}
A_{n}^{N=4}\ \equiv\
A_{n}^{[1]} + 4A_{n}^{[1/2]}+3 A_{n}^{[0]}\,.
\end{equation}
The contributions from these three multiplets 
are not independent but satisfy 
\begin{equation}
A_{n}^{N=1\ {\rm vector}}\ \equiv\ A_{n}^{N=4} -
3A_{n}^{N=1\ {\rm chiral}} \,.
\end{equation}
Thus, provided the $N=4$ amplitude is known, one must only calculate one
of the two possibilities for $N=1$. The amplitude
$A^{N=4}(1^-,2^-,3^-,4^+,5^+,6^+)$ has been calculated~\cite{BDDKb}, so in
this letter we choose to examine the $N=1$ chiral matter multiplet contribution.


\section{The Six-point Amplitude $A^{N=1,{\rm chiral}}(1^-,2^-,3^-,4^+,5^+,6^+)$ }

For a six-point Yang-Mills amplitude there are a relatively small
number of independent colour-ordered helicity configurations.  The
non-vanishing supersymmetric amplitudes are either MHV, the conjugate of
MHV (googly), or have three negative and three positive helicities.  

The MHV amplitudes are rather special cases and indeed 
the holomorphic anomaly of the
three particle cuts of 
$A^{N=1,{\rm chiral}}(1^-,2^-,3^+,4^+,5^+,6^+)$
is zero and is a rather uninteresting case. 
Consequently, 
we consider an amplitude with three negative helicities
which has a richer
structure.  There are three possible such colour-ordered configurations: 
$A(---+++)$, $A(--+-++)$ and $A(-+-+-+)$. We shall consider the effect of
the holomorphic anomaly on the first of these. The amplitude $A(---+++)$ has
not been published previously.  (It can be viewed as a component of
$A_n^{[1]}$ and $A_n^{[1/2]}$, and thereby contributes to a six-gluon QCD 
amplitude required for the next-to-leading order production of four jets at hadron
colliders.)

This amplitude is fairly simple in that it contains no box
integral functions, but  only  $\Lz$ and
$\Kz$\ functions. The amplitude is
\begin{equation}
\eqalign{ A_6^{N=1\ {\rm chiral}}( & 1^-,2^-,3^-,4^+,5^+,6^+) = a_1 K_0[s_{61}] +a_2
K_0[s_{34}] \cr 
& - {i\over 2} \Biggl[ b_1 { L_0 [s_{345}/ s_{61} ] \over s_{61} } +b_2{
L_0 [ s_{234} /s_{34} ] \over s_{34}} +b_3{ L_0 [ s_{234}/s_{61} ]
\over s_{61}} +b_4{ L_0 [ s_{345} /s_{34} ] \over s_{34} } 
\Biggr] \cr}
\label{specificform}
\end{equation}
where the coefficients are
\begin{equation}
a_1 =a_2 = {1\over2} \Atree_6(1^-,2^-,3^-,4^+,5^+,6^+),
\end{equation}
and
\begin{equation}
\eqalign{ b_1 =& {  \la 6 | \Slash P  | 3 \ra ^2\la 6^+ | (
\Slash2\Slash P\Slash P-\Slash P\Slash2\Slash P ) | 3^+ \ra \over
\la 2 | \Slash P | 5 \ra \spb6.1\spb1.2\spa3.4\spa4.5 P^2 } \,, 
\hskip 1.0 truecm P=P_{345}\equiv k_3+k_4+k_5, \cr 
b_2 =& { \la 4 | \Slash
P  | 1 \ra ^2\la 4^+ | ( \Slash P\Slash2\Slash P-\Slash2\Slash
P\Slash P ) | 1^+ \ra \over \la 2 | \Slash P | 5 \ra
\spb2.3\spb3.4\spa5.6\spa6.1 P^2 } \,, \hskip 1.0 truecm
P=P_{234}\equiv k_2+k_3+k_4, \cr 
b_3 =& { \la 4 | \Slash P | 1 \ra^2
\la 4^+ | ( \Slash P\Slash P\Slash5-\Slash P\Slash5\Slash P ) | 1^+ \ra 
\over \la 2 | \Slash P  | 5 \ra
\spb2.3\spb3.4\spa5.6\spa6.1 P^2 } \,, \hskip 1.0 truecm P=P_{234}, \cr
b_4 =& {  \la 6 | \Slash P  | 3 \ra ^2\la 6^+ | ( \Slash P \Slash5
\Slash P-\Slash P\Slash P\Slash 5 ) | 3^+ \ra \over  \la 2 |
\Slash P  | 5 \ra \spb6.1\spb1.2\spa3.4\spa4.5 P^2 } \,, \hskip 1.0
truecm P=P_{345}. \cr}
\end{equation}
The six-point tree amplitudes were calculated in ref.~\cite{MPX}. 
The amplitude has an overall factor in dimensional regularisation of 
$(\mu^2)^{\eps} c_\Gamma$, where
\begin{equation}
c_\Gamma\ =\ {1 \over (4 \pi)^{2-\e}}
{\Gamma(1+\e)\Gamma^2(1-\e)\over\Gamma(1-2\e)} \,,
\label{cGamma}
\end{equation}
which we do not write explicitly.  
We define 
$s_{ij} \equiv \spb{i}.j\spa{j}.i$,
$s_{ijk} \equiv P^2_{ijk}
\equiv \spb{i}.j\spa{j}.i + \spb{j}.k\spa{k}.j + \spb{k}.i\spa{i}.k\equiv(k_i+k_j+k_k)^2$
and 
$\la a | \Slash b | c \ra \equiv \la a^+ | \Slash b   | c^+\ra
\equiv [ab]\la bc\ra$,
where $\spa{i}.j$ and $\spb{i}.j$ are the usual spinor
helicity inner products~\cite{SpinorHelicity}.
The $\Lz$ and $\Kz$ functions are defined by
\begin{equation}
\eqalign{ \Kz [s] \ =&\  {1 \over \eps(1-2\eps)  } ({-s})^{-\eps}
=  {1\over\eps} - \ln(-s) + 2  + {\cal O}(\e)  \; ,  \cr 
\Lz [ r ] \ =&\ { {\rm ln} (r) \over 1-r } + {\cal O}(\e)\;  . \cr}
\end{equation}
The function $\Kz[s]$ is simply proportional to the scalar bubble
function. The function $\Lz[r]$ has several representations; it can be
expressed as a linear combination of bubble functions or 
as a Feynman parameter integral for a two-mass triangle integral~\cite{BDDKb}.

This amplitude was constructed by calculating the three-particle cuts
together with an analysis of the infra-red poles. We shall be
revisiting the three-particle cuts when we consider the action of
the holomorphic anomaly. We exhibit the limit where legs 1 and 2 
become collinear to illustrate the consistency of the amplitude at this 
infra-red pole. 
The other limits are analogous, although not identical.


\noindent
{\bf Collinear limit $1-2$}

The collinear limits of the (colour-ordered) one-loop partial
amplitudes have the following form:
\begin{equation}
\eqalign{
A_{n}^{\rm loop}\ \mathop{\longrightarrow}^{a \parallel b}\
\sum_{\lambda=\pm} & \biggl(
  \Split^{\rm tree}_{-\lambda}(a^{\lambda_a},b^{\lambda_b})\,
      A_{n-1}^{\rm loop}(\ldots(a+b)^\lambda\ldots)
\cr
&  +\Split^{\rm loop}_{-\lambda}(a^{\lambda_a},b^{\lambda_b})\,
      A_{n-1}^{\rm tree}(\ldots(a+b)^\lambda\ldots) \biggr) ,
\cr}
\end{equation}
where $k_a \to z k_P$, $k_b \to (1-z)k_P$, and $\lambda$ is the helicity
of the state $P$. The splitting amplitudes are universal 
and may be derived, for example, from the five-gluon
amplitudes~\cite{FiveGluon,BDDKa}.

For supersymmetric theories the loop splitting amplitudes
$\Split^{\rm loop}_{-\lambda}(a^{\lambda_a},b^{\lambda_b})$ are proportional to the tree splitting
amplitudes,
\begin{equation}
 \Split^{\rm loop}_{-\lambda}(a^{\lambda_a},b^{\lambda_b})
   \ =\ {c_\Gamma }
  \times \Split^{\rm tree}_{-\lambda}(a^{\lambda_a},b^{\lambda_b})
  \times r_S^{\rm SUSY}(z,s_{ab}).
\end{equation}
For the $N=1$ matter multiplet (but not the vector) the collinear limit is simplified 
since
\begin{equation}
r_S^{N=1\ {\rm chiral}}(z,s_{ab}) = 0
\end{equation}
for collinear gluons.  This reduces the collinear condition to
\begin{equation}
\eqalign{
A_{n}^{\rm loop}\ \mathop{\longrightarrow}^{a \parallel b}\
\sum_{\lambda=\pm}
  \Split^{\rm tree}_{-\lambda}(a^{\lambda_a},b^{\lambda_b})\,
&
A_{n-1}^{\rm loop}(\ldots(a+b)^\lambda\ldots) \,.
\cr}
\end{equation}

The `target' of the $1-2$ collinear limit will be the five-point 
$N=1$ matter amplitude~\cite{FiveGluon},
\begin{equation}\eqalign{
A_5^{N=1\ {\rm chiral} }(1^-,3^-,4^+,5^+,6^+) =&  { \Atree_5 \over 2} \biggl[   K_0[s_{34}]
+ K_0[s_{61}] 
\cr & \hskip0.8cm
-  {  L_0[ s_{345}/s_{34} ]  \over s_{34} }
 { \tr_+ [1 3 5 (1+6) ] - \tr_+[13(1+6)5] \over s_{13} } \biggr] \,. \cr}
\end{equation}
In the limit $k_1\rightarrow z k_1, k_2\rightarrow (1-z) k_1$, the
coefficients $b_2$ and $b_3$ are non-singular, where 
$\tr_+[ijkl] = \spb{i}.j\spa{j}.k\spb{k}.l\spa{l}.i $.  Using the collinear
properties of tree amplitudes, the coefficient $a_2$ becomes
\begin{equation}
a_2 = { \Atree_6 \over 2 }\ \longrightarrow\ 
S^{--}_{+} { \Atree_5(1,3,4,5,6) \over 2 } \,,
\end{equation}
where $S^{--}_{+} \equiv \Split^{\rm tree}_{+}(1^-,2^-)$.
So the combination $a_2 K_0[s_{34}]$ behaves as,
\begin{equation}
a_2 K_0[s_{34}] 
\ \longrightarrow\ S^{--}_+ { \Atree_5(1,3,4,5,6)\over 2 } K_0[s_{34}] \,,
\end{equation}
which is one of the required terms.

The integral function multiplying $b_4$ trivially goes
to the function $\Lz[s_{345}/s_{34}]$.  The coefficient $b_4$ approaches  
\begin{equation}
\eqalign{ b_4\ \longrightarrow\ & 
{ -1 \over \sqrt{z(1-z)} \spb1.2 }
\times {  \la 6 | \Slash 1  | 3 \ra ^2\la 6^+ | ( (\Slash{1}+\Slash{6}) \Slash
5(\Slash{1}+\Slash{6})-  (\Slash{1}+\Slash{6})(\Slash{1}+\Slash{6})\Slash 5 )| 3^+ \ra \over  \la 1 |
\Slash 6  | 5 \ra \spb6.1\spa3.4\spa4.5 s_{61} } \cr 
=\ & S^{--}_+
\times { \spb6.1^2 \spa1.3^2 \over \spb1.6\spa6.5
\spb6.1\spa3.4\spa4.5 s_{61} } \times 
\la 6^+ | [  \Slash{1} \Slash{5} (\Slash{1}+\Slash{6}) 
         - \Slash{1} \Slash{6} \Slash 5 ] | 3^+ \ra \cr 
= & - S^{--}_+ { \Atree_5 \over i } 
  \times { \tr_{+} [ 13 ((6+1)5-5(6+1)) ] \over s_{13} } \,, \cr}
\end{equation}
after some rearrangement, which is as required.

\noindent The remaining two integrals must combine in the
collinear limit.  An identity similar to eq.~(III.10) of
ref.~\cite{BDDKb} can be used,
\begin{equation}\eqalign{
(1-z)  {\Lz[ s_{612} / s_{61} ]\over s_{61}}
+ z { \Kz[ s_{61} ] \over s_{61} }
\ &\longrightarrow\  { \Kz[ s_{61} ] \over s_{61} } \,.
\cr}
\label{CombineTwo}
\end{equation}
The limits of the combination $a_1 s_{61}$ and of $b_1$ are both
proportional to the five-point tree amplitude:
\begin{equation}
a_1 s_{61}
\longrightarrow   S^{--}_+ \times { \Atree_5(1,3,4,5,6) \over 2} \times s_{61}  \times z,
\end{equation}
and
\begin{equation}
\eqalign{ b_1\ \longrightarrow\ & { - (1-z) \over \sqrt{z(1-z)}
\spb1.2   }  \times {  \la 6 |\Slash 1  | 3 \ra ^2\la 6^+ | (
\Slash 1\Slash 6\Slash 1-\Slash 1\Slash 1( \Slash{6}+\Slash{1})  ) | 3^+ \ra
\over  \la 1 | \Slash 6 | 5 \ra \spb6.1\spa3.4\spa4.5 s_{61} } \cr
=\   & S_+^{--} \times (1-z) \times { ( \spb6.1\spa1.3  )^2 s_{61}
\la 6 |  1  | 3 \ra \over \spb1.6 \spa6.5 \spb6.1\spa3.4\spa4.5
s_{61} } \cr =\ & - S_+^{--} \times (1-z) \times {\Atree_5 \over i }
\times s_{61} \,. \cr}
\end{equation}
Thus, using \eqn{CombineTwo}, we have
\begin{equation}
a_1 K_0[s_{61}] - {i\over2} b_1 { L_0 [s_{612}/s_{61} ] \over s_{61} }
 \longrightarrow S^{--}_+ \times { \Atree_5(1,3,4,5,6) \over 2}
\times \Kz[s_{61}] \,.
\end{equation}
Adding all the pieces together, we find that the amplitude has the 
correct collinear limit.

The other collinear limits are similar. The limits $2-3$,
$4-5$ and $5-6$ are related by symmetry; they follow from the 
$1-2$ limit by relabelling and conjugation. The $6-1$ and $3-4$ 
limits are different but can be shown to have the correct limit 
in an analogous manner.  The amplitude also has the correct 
multi-particle poles~\cite{Bern:1995ix} when $s_{234} \to 0$ 
or $s_{345} \to 0$. 



\section{Holomorphic Anomaly of the Unitary Cuts}

The amplitude we are considering has three potential three-particle
cuts: $s_{123}>0$, $s_{234}>0$ and $s_{345}>0$.  The first of these
vanishes identically for $N=1$ matter: 
$\Atree_5(\ell_1^{h_1},1^-,2^-,3^-,\ell_2^{h_2})=0$
unless $h_1 = h_2 = 1$, which requires the states crossing the cut
to be gluons, not fermions or scalars.
The two non-vanishing cuts are not independent but may be
obtained from one another by the symmetry $1 \lr 3, 4 \lr 6$.

In order to examine the holomorphic anomaly, we compute the
action of $F_{561}$ on the cut $C_{561}$ (which is equal to $C_{234}$). 
The cut for $s_{561}> 0$ (the imaginary part, or $1/2$ the
discontinuity) is defined as
\begin{equation}
C_{561} = {i \over 2} \int \dlips \sum_{h\in \{-1/2,0,1/2\}} \Atree_5(\ell_1^h,5^+,6^+,1^-,\ell_2^{-h})
\Atree_5((-\ell_2)^h, 2^-,3^-,4^+,(-\ell_1)^{-h}) \,,
\end{equation}
where $\ell_1+\ell_2 = P_{234} \equiv P$.
Writing out all amplitudes in this expression and summing over the
supersymmetric multiplet we obtain
\begin{equation}
C_{561} = {i \over 2}\int \dlips \frac{\spa1.{\ell_1}^2 \spa1.{\ell_2}^2}
{\spa5.6\spa6.1\spa1.{\ell_2}\spa{\ell_2}.{\ell_1}\spa{\ell_1}.5} 
\times 
\frac{\spb4.{\ell_1}^2 \spb4.{\ell_2}^2}
{\spb2.3\spb3.4\spb4.{\ell_1}\spb{\ell_1}.{\ell_2}\spb{\ell_2}.2} 
\times \rho_{N=1} \,.
\end{equation}
The factor $\rho_{N=1}$ may be obtained using 
supersymmetric Ward identities~\cite{SWI},
giving  
\begin{equation}
\rho_{N=1} = 
\frac{\langle 4 | \Slash{P} | 1 \rangle^2}{\spa1.{\ell_1}\spb{\ell_1}.4\spa1.{\ell_2}
\spb{\ell_2}.4} \; . 
\end{equation}
Simplifying the expression, we can write $C_{561}$ in a compact form
\begin{equation}\begin{split}
C_{561} & = i{K \over 2} \int \dlips
        \frac{\spb4.{\ell_2}\spa1.{\ell_1}}{\spb2.{\ell_2}\spa5.{\ell_1}} \,,
\end{split}\end{equation}
where we define $K$ as
\begin{equation}
K = \frac{\langle 4 | \Slash{P}_{234} | 1 \rangle^2}{\spb2.3
\spb3.4\spa5.6\spa6.1 s_{234} } \,.
\end{equation}
Next we act with the collinear operator $[F_{561},\eta]$ 
on this expression. It is clear that we will only pick up the contribution from
the term with ${\partial}/{\partial \tilde \lambda_{5\dot a}}$,
so that
\begin{equation}\begin{split}
[F_{561},\eta]C_{561} = { iK\over 2}
        \int \dlips
        \frac{\spb4.{\ell_2}\spa1.{\ell_1}}{\spb2.{\ell_2}}\spa6.1\left 
[\frac{\partial}{\partial \tilde \lambda_5},
        \eta\right]\frac{1}{\spa5.{\ell_1}} \,.
\end{split}\end{equation}
The parametrization of the Lorentz-invariant phase-space measure $\dlips$ 
is the same as that
employed in~\cite{Cachazo:2004kj,Bena:2004xu}, {\it i.e.,}
\begin{equation}
\eqalign{
\int \dlips(\bullet) &\equiv \int d^4 \ell_1\
\delta^{(+)}(\ell_1^2)\int d^4 \ell_2
\delta^{(+)}(\ell_2^2)\delta^{(4)}(\ell_1+\ell_2-P)(\bullet)
\cr
&=\int_0^\infty tdt \int \langle\lambda_{\ell_1},d\lambda_{\ell_1}\rangle [\tilde
\lambda_{\ell_1},d\tilde\lambda_{\ell_1}]
\int d^4 \ell_2
\delta^{(+)}(\ell_2^2)\delta^{(4)}(\ell_1+\ell_2-P)
(\bullet),
\cr}
\end{equation}
and we change coordinates, $\lambda \rightarrow \lambda'$ 
and $\tilde \lambda \rightarrow t \tilde\lambda'$, then drop the primes. 
Hence the integral becomes
\begin{equation}\begin{split}
[F_{561},\eta]C_{561} &= i{K\over2} \int_0^\infty tdt \int \langle
\lambda_{\ell_1},d\lambda_{\ell_1}\rangle [\tilde
\lambda_{\ell_1},d\tilde\lambda_{\ell_1}]\\&\int d^4 \ell_2
\delta^{(+)}(\ell_2^2)\delta^{(4)}(\ell_1+\ell_2-P) \frac{\spb4.{
\ell_2}\spa1.{\ell_1}\spa6.1}{\spb2.{\ell_2}}
\left[\frac{\partial}{\partial \tilde \lambda_5},
\eta\right]\frac{1}{\spa5.{\ell_1}} \,.
\end{split}\end{equation}

We follow the prescription of ref.~\cite{Cachazo:2004dr} and
use the identity,
\begin{equation}
\left[\frac{\partial}{\partial \tilde \lambda_5},\eta\right]
\frac{1}{\spa{\ell_1}.{5}} =
-\left[\frac{\partial}{\partial \tilde
\lambda_{\ell_1}},\eta\right] \frac{1}{\spa{\ell_1}.5} \,,
\end{equation}
which can be rewritten using the Schouten identity
$\spb{A}.B\spb{C}.D = \spb{A}.C\spb{B}.D - \spb{A}.D\spb{B}.C$,
so that
\begin{equation}
[\tilde \lambda_{\ell_1},d \tilde \lambda_{\ell_1}]
\left[\frac{\partial}{\partial \tilde
\lambda_{\ell_1}},\eta\right] =
\left[\tilde\lambda_{\ell_1},\frac{\partial}{\partial \tilde
\lambda_{\ell_1}}\right] [d\tilde\lambda_{\ell_1},\eta] -
[\tilde\lambda_{\ell_1},\eta]\left[d\tilde\lambda_{\ell_1},\frac{\partial}{\partial
\tilde \lambda_{\ell_1}}\right] \,,
\end{equation}
where the first term does not contribute to the integral. Hence 
inside the integral we can rewrite
\begin{equation}
[\tilde \lambda_{\ell_1},d \tilde \lambda_{\ell_1}]
\left[\frac{\partial}{\partial \tilde \lambda_5},\eta\right]
\frac{1}{\spa{\ell_1}.5} =
[\tilde\lambda_{\ell_1},\eta]\left[d\tilde\lambda_{\ell_1},\frac{\partial}{\partial
\tilde \lambda_{\ell_1}}\right] \frac{1}{\spa{\ell_1}.5}
=[\tilde\lambda_{\ell_1},\eta]
2\pi \bar\delta(\langle \lambda_{\ell_1},\lambda_5 \rangle)\; , 
\end{equation}
and the integral becomes
\begin{equation}\begin{split}
[F_{561},\eta]C_{561}& =-i\pi K\int_0^\infty tdt \int \la
\lambda_{\ell_1},d\lambda_{\ell_1}\rangle \\
&\int d^4 \ell_2
\delta^{(+)}(\ell_2^2)\delta^{(4)}(\ell_1+\ell_2-P) \frac{\spb{4}.{\ell_2}
\spa{1}.{\ell_1}\spa6.1
[\tilde\lambda_{\ell_1},\eta]}{\spb2.{\ell_2}} \bar
\delta(\langle \lambda_{\ell_1},\lambda_5 \rangle) \,.
\end{split}\end{equation}
The $\delta$ function in $\langle \lambda_{\ell_1},\lambda_5 \rangle$ 
reduces the integral to
\begin{equation}
[F_{561},\eta]C_{561} = -i\pi K \int_0^\infty tdt \int d^4 \ell_2
\delta^{(+)}(\ell_2^2)\delta^{(4)}(\ell_1+\ell_2-P) \frac{\spb4.{
\ell_2}\spa{\ell_2}.a\spa6.1\spa1.5 [5,\eta]}{\spb2.{\ell_2}\spa{\ell_2}.a}
 \,.
\end{equation}
We have introduced a factor of $\spa{\ell_2}.a/\spa{\ell_2}.a$, 
which makes applying the $\delta$ function 
in $\ell_2$ more transparent. 
Doing the integral in $\ell_2$ using
$\delta^{(4)}(\ell_1+\ell_2-P)$ we end up with
\begin{equation}
[F_{561},\eta]C_{561} =-i\pi  K\int_0^\infty tdt\delta^{(+)}(\ell_2^2)
\frac{\spb4.{\ell_2}\spa{\ell_2}.a \spa6.1 \spa1.5 [5,\eta]}{ 
\spb2.{\ell_2}\spa{\ell_2}.a} \,,
\end{equation}
where now $\ell_2^\mu = P^\mu - t k_5^\mu$, and hence
$\ell_2^2 = P^2 - 2tk_5\cdot P$, where $t = {{P^2} \over
{2k_5\cdot P}}$. 
Doing the $t$-integral yields
\begin{equation}
\eqalign{
[F_{561},\eta]C_{561} 
&
= i\pi  K 
{     \spa6.1\spa1.5  [5,\eta] P^2
\over 
( 2k_5\cdot P )^2 }  
{
{ (2k_5\cdot P) \la 4 |\Slash{P}|a\rangle - P^2\la4 |\Slash 5 |a\rangle }
\over 
{ (2k_5\cdot P) \la 2 |\Slash{P}|a\rangle - P^2\la2 |\Slash 5 |a\rangle }
}
\cr
&
=
i\pi \frac{\langle 4 |\Slash P | 1\rangle^2 }{\spb2.3\spb3.4\spa5.6}
 {{\spa1.5  [5,\eta]} \over {(2k_5\cdot P)^2}}
{ \la 4 |\Slash P|5 \ra \over \la 2 |\Slash P|5 \ra  } \,,
\label{tempint}
\cr}
\end{equation}
after reinstating the definition of $K$ and choosing $a=3$, for example.

From the optical theorem, the cut $C_{561}$ is equal to 
the imaginary part of the amplitude in the kinematic region 
$s_{561} > 0$~\cite{Cutting}. 
For our amplitude~(\ref{specificform}), using 
${\rm Im} \ln(-s)|_{s > 0} = - \pi$,  
this is
\begin{equation}
-{1\over \pi} {\rm Im} A_{s_{561} > 0} = -{i\over2} \biggl[
\frac{b_3}{2k_5\cdot P} - \frac{b_2}{2k_2\cdot P} \biggr] \,.
\label{Imspecific}
\end{equation}
Operating on \eqn{Imspecific} with the collinear operator 
$[F_{561},\eta]$ we have
\begin{equation}
[F_{561},\eta] \left( -{1\over \pi} {\rm Im} A_{s_{561} > 0} \right)
=  -{i\over2} \biggl[ \frac{[F_{561},\eta](b_3)} {2k_5\cdot P} 
- \frac{[F_{561},\eta](b_2)} {2k_2\cdot P}-
\frac{b_3[F_{561},\eta](2k_5\cdot P)}{(2k_5\cdot P)^2} \biggr] \,. 
\label{eqIm}
\end{equation}
Using the solutions for $b_2$ and $b_3$ we have
\begin{equation}
b_3 = \frac{\langle 4 | \Slash P | 1 \rangle^2}{\langle 2 |\Slash
P| 5\rangle}\frac{\langle 4^+ | \Slash P\Slash P \Slash 5 - \Slash
P \Slash 5 \Slash P | 1^+ \rangle}{\spb2.3\spb3.4\spa5.6\spa6.1
 P^2} = \frac{K}{\langle 2 |\Slash
P| 5\rangle}\langle 4^+ | \Slash P\Slash P \Slash 5
- \Slash P \Slash 5 \Slash P | 1^+ \rangle 
\equiv K' \hat b_3 \,,
\label{b3}
\end{equation}
where 
\begin{equation}
K' \equiv \frac{K}{\langle 2 |\Slash{P}| 5\rangle}
\label{Kprimedef}
\end{equation}
is annihilated by $F_{561}$, and where
$$
\hat b_3
\equiv  2P^2 \la 4 | \Slash 5 | 1\ra  -(2k_5\cdot P) \la 4 | \Slash P | 1\ra \,.
$$
Also,
\begin{equation}
 b_2 = { \la 4 | \Slash P  | 1 \ra
^2\la 4^+ | ( \Slash P\Slash2\Slash P-\Slash2\Slash P\Slash P ) |
1^+ \ra \over \la 2 | \Slash P | 5 \ra
\spb2.3\spb3.4\spa5.6\spa6.1 P^2 } = \frac{K}{\langle 2 |\Slash
P| 5\rangle} \la 4^+ | ( \Slash
P\Slash2\Slash P-\Slash2\Slash P\Slash P ) | 1^+ \ra \equiv K' \hat
b_2 \,,
\end{equation}
where
\begin{equation}
\hat b_2 \equiv - 2 P^2 \la 4 | \Slash 2 | 1 \ra
                   + (2 k_2\cdot P) \la 4 | \Slash P | 1 \ra \,.
\end{equation}
Using
$
[F_{561},\eta](2k_5\cdot P) 
= \langle \eta | \Slash{P} | 5\rangle \langle 16\rangle
$,
we have
\begin{equation}\begin{split}
[F_{561},\eta] \hat b_3 &= 2P^2 \langle 51 \rangle [F_{561},\eta]
[45] - \langle 4 | \Slash{P} | 1 \rangle [F_{561},\eta] (2k_5 \cdot P )\\
 & = - 2P^2 [\eta, 4] \spa1.6\spa5.1 - \langle
 \eta | \Slash{P} | 5 \rangle \spa1.6 \langle 4 | \Slash{P} | 1 \rangle \,,
\label{Fhatb3}
\end{split}\end{equation}
and
\begin{equation}\begin{split}
[F_{561},\eta] \hat b_2 = 0.
\label{Fhatb2}
\end{split}\end{equation}
Inserting \eqns{Fhatb3}{Fhatb2} into eq.~(\ref{eqIm}), we find,
\begin{equation}\begin{split}
-{1\over\pi} [F_{561},\eta] {\rm Im } A 
&= - {i\over2} \frac{K'}{2k_5\cdot P} \Big( - 2P^2
[\eta, 4] \spa1.6 \spa5.1 - \langle \eta |
\Slash P | 5\rangle \spa1.6 \langle 4 |\Slash P | 1
\rangle\\
&\hskip2.5cm 
 - \frac{2P^2 \langle 4 |\Slash 5 | 1 \rangle \langle
\eta |\Slash P | 5\rangle\spa1.6}{2k_5 \cdot P} +
\langle \eta |\Slash P | 5\rangle\spa1.6\langle
4|\Slash P| 1\rangle \Big) \\
& = - {i\over2} \frac{2K' P^2\langle
16\rangle\langle 15\rangle}{(2 k_5 \cdot P)^2} \Big [[\eta, 4]
(2k_5\cdot P) + \langle \eta |\Slash P | 5 \rangle\spb4.5\Big] \,.
\end{split}\end{equation}
Combining $[\eta, 4][P, 5] - [\eta, P][45] = [\eta, 5][P, 4]$ using the
Schouten identity, we obtain
\begin{equation}\begin{split}
- {1\over\pi} [F_{561},\eta] {\rm Im } A 
& = - i \frac{K' P^2\spa1.6\spa1.5\langle 5, P \rangle}{(2 k_5 \cdot P)^2}
\Big [[\eta,5][P, 4]\Big]
= - i \frac{\langle 4 |\Slash P | 1\rangle^2   \spa1.5}{\spb2.3\spb3.4\spa5.6 }
{[5,\eta]\over(2k_5\cdot P)^2}{\langle 4 |\Slash P |
5\rangle \over \langle 2|\Slash P|5\rangle} \,,
\end{split}\end{equation}
which matches the expression in eq.~(\ref{tempint}). Thus we have shown
that the holomorphic anomaly of the unitarity cuts correctly reproduces
the action of $F_{ijk}$ upon the imaginary part of the amplitude.



\section{Reconstructing Amplitudes from Differential Equations}

In $N=4$ one-loop amplitudes, appropriate collinear operators $F_{ijk}$
annihilate the coefficients of the scalar box integral functions which span the
amplitude~\cite{Cachazo:2004dr}.  This has the implication that the
coefficients may be reconstructed by solving algebraic equations
resulting from the action of the $F_{ijk}$ operator upon the
cuts equation.  For $N=1$ we have a more delicate situation as the
collinear operator $F_{ijk}$ in this case acts non-trivially on the
coefficients $b_i$ in the amplitude. This means that to
reconstruct the amplitude we will generally have to solve differential equations
for the coefficients $b_i$.  In this section we explore the
possibility of reconstructing the amplitude using the holomorphic
anomaly of the cuts.  In general $N=1$ amplitudes contain integral
functions derived from box, triangle and bubble integrals. As for the
$N=4$ case, we expect that the appropriate $F_{ijk}$ operators should
annihilate the coefficients of the box integral functions. 
However, $F_{ijk}$ need not annihilate the coefficients
of bubble and triangle functions.  Instead, the action of $F_{ijk}$ produces
differential equations which these coefficients must satisfy.

To clarify the situation, consider the amplitude 
$A^{N=1\ \rm chiral}(1^-,2^-,3^-,4^+,5^+,6^+)$ 
which contains only triangle and bubble integrals.
Consider the action of $F_{561}$  on the $C_{561}$ cutting equation,
\begin{equation}
\eqalign{
[ F_{561}, \eta]{\rm Im} A_{s_{561} >0}   =&\  
[ F_{561}, \eta] C_{561}  \,.
\cr}
\end{equation}
Expanding the amplitude according to eq.~(\ref{generalform}) and keeping only those
coefficients which have non-vanishing cuts in this channel, 
namely $b_2$ and $b_3$ in eq.~(\ref{specificform}), we have
\begin{equation}
\eqalign{
{i\pi \over2} [ F_{561}, \eta] \left(  { b_3 \over { 2 k_5 \cdot P } } 
-  { b_2 \over { 2 k_2 \cdot P } } \right)=&\ 
[ F_{561}, \eta] C_{561} \,.
\cr}
\label{Fb3b2}
\end{equation}
The right-hand side of this equation is a rational function of
$\lambda_i$ and $\tilde\lambda_j$, determined  {\it via} the
holomorphic anomaly to be the expression given in eq.~(\ref{tempint}).
In \eqn{Fb3b2} the functions multiplying the $b_i$ are rational functions 
--- in contrast to the $N=4$ situation where logarithms appear.  
Although the left-hand side is required to be rational this
does not imply that $F_{ijk}$ annihilate the $b_i$.  
The $b_i$ must satisfy the linear differential equation
\begin{equation}
 {i\pi \over2} \biggl[ 
\frac{[F_{561},\eta]b_3}{ 2k_5\cdot P }
- \frac{b_3[F_{561},\eta](2k_5\cdot P) }{(2k_5\cdot P)^2}
- \frac{[F_{561},\eta]b_2}{ 2k_2\cdot P }
\biggr]
= [ F_{561}, \eta] C_{561} \,. 
\end{equation}

We can also act with the operator 
\begin{equation}
\la \bar F_{ijk} , \bar\eta \ra=
\spb{i}.j \biggl\langle { \partial\over \partial\lambda_k},\bar\eta\biggr\rangle
+\spb{j}.k  \biggl\langle{ \partial\over
\partial\lambda_i},\bar\eta\biggr\rangle +\spb{k}.i \biggl\langle{
\partial\over
\partial\lambda_j},\bar\eta\biggr\rangle \,,
\end{equation}
which produces an ``anti-holomorphic anomaly''
upon the same cut to
yield
\begin{equation}
 {i\pi \over2} \biggl[ 
\frac{\la\bar
F_{234},\bar\eta\ra b_2}{ 2k_2\cdot P }
- \frac{b_2 \la\bar F_{234},\bar
\eta\ra (2k_2\cdot P) }{(2k_2\cdot P)^2}
- \frac{\la\bar
F_{234},\bar\eta \ra b_3}{ 2k_5\cdot P } 
\biggr] 
= 
\la \bar F_{234}, \bar\eta\ra C_{561} \,. 
\end{equation}
As a function of $\tilde\lambda_5$, $\tilde\lambda_6$,and
$\tilde\lambda_1$, we find explicitly that $[ F_{561}, \eta]
C_{561}$ is a function of $\tilde\lambda_5$ only. Similarly $\la
\bar F_{234}, \bar\eta\ra C_{561}$ is a function of $\lambda_2$
only.  
The coefficients $b_2$ and $b_3$ are related by the
symmetry of the amplitude to satisfy $b_2(123456)=\bar
b_3(456123)$.  
Also note that $\la \bar F_{234}, \bar\eta\ra[ F_{561},
\eta] C_{561}=0$. 
This motivates us to separate the equations, 
by assuming that $\la \bar F_{234},
\bar\eta\ra b_3=0$ and $[ F_{561}, \eta] b_2=0$, to obtain the
equation for $b_3$,
\begin{equation}
{i\pi \over2} \biggl[ 
\frac{[F_{561},\eta]b_3}{ 2k_5\cdot P }
- \frac{b_3[F_{561},\eta](2k_5\cdot P) }{(2k_5\cdot P)^2}
\biggr]
= 
[ F_{561}, \eta] C_{561}
\label{diffE1}
\end{equation}
(with the equation for $b_2$ obtained by relabelling). To solve
this equation, it is convenient to define
\begin{equation}
b_3 = K'  \hat b_3
\end{equation}
as in eq.~(\ref{b3}).  Note that $K'$ is independent of $\tilde\lambda_i$,
$i=5,6,1$. Since \eqn{diffE1} is independent of $\tilde\lambda_i$,
$i=6,1$, we deduce that $b_3$ depends only on $\tilde\lambda_5$.
The right-hand side of eq.~(\ref{diffE1}), from \eqn{tempint},
\begin{equation}
{ [ F_{561}, \eta] C_{561} \over K' } = i \pi \frac{P^2\langle
16\rangle\langle 15\rangle\langle 5, P \rangle}{(2 k_5 \cdot P)^2}
\Big [[\eta, 5][P, 4]\Big] \,,
\end{equation}
is of the form $[X,5]$.  So we make a trial solution for $\hat b_3$
\begin{equation}
\hat b_3 = [5,{\cal C}] \,, 
\end{equation}
which implies
\begin{equation}
\eqalign{
\frac{[F_{561},\eta]\hat b_3}{(2k_5\cdot P)} -
\frac{\hat b_3[F_{561},\eta](2k_5\cdot P) }{(2k_5\cdot P)^2}
& =
- \frac{[5,P]\langle P,5 \rangle[\eta, {\cal C}]\spa6.1}{(2k_5\cdot P)^2} + 
\frac{[5, {\cal C}]\spa6.1 [\eta, P]\langle P,5
\rangle }{(2k_5\cdot P)^2} 
\cr
& = \frac{\spa1.6 \langle 5,P
\rangle }{(2k_5\cdot P)^2}\Big[[\eta, 5][P,{\cal C}] \Big] \,.
\cr}
\end{equation}
Thus eq.~(\ref{diffE1}) is solved by 
\begin{equation}
{\cal C}_{\dot a} = 2P^2 \langle 15 \rangle \tilde \lambda_{4\dot a} \,,
\end{equation}
giving
\begin{equation}
\hat b_3 = 2P^2 \la 4 | \Slash{5} | 1 \ra 
\end{equation}
as a specific solution to eq.~(\ref{diffE1}). 
However, this solution is not unique, as 
\begin{equation}
\hat b_3 =2P^2 \la 4 | \Slash{5} | 1 \ra  + (2k_5 \cdot P) \times A  
\end{equation} 
is also a solution, for any rational function $A$ not 
involving $\tilde\lambda_i$, $i=5,6,1$. 
To also satisfy\break  $\la\bar F_{234} , \bar\eta\ra (
b_3/(2k_5\cdot P) )=0$, we must have,
\begin{equation}
\la\bar F_{234} , \bar\eta\ra A = 0.
\label{Fannih}
\end{equation}
This relation is not sufficient to fix $A$.  Indeed, any function of
$P_{a\dot a}=\sum_{i=5,6,1} (\lambda_i)_a (\tilde\lambda_i)_{\dot a}$
will satisfy \eqn{Fannih}. 
We have used the action of all $F_{ijk}$ functions
which give rational functions acting upon the cut.  The information
in other cut channels is equivalent to this cut by relabelling. Thus
we are led to conclude that the action of the $F_{ijk}$ operators upon
the cuts does not uniquely fix the coefficients without the input of
further information. In some sense, acting upon the cut with
differential operators is destroying information which must be covered
by examining boundary conditions or other constraints. Examples of the 
constraints that $\hat b_3$ must satisfy are: 
dimensionality, spinor weight, collinear limits, multi-particle poles,
{\it etc.} 
For example, the coefficient $\hat b_3$ must have dimension 2 and 
the spinor weight of $+1$ with respect to leg $4$, $-1$ with respect to 
leg $1$, and $0$ for other legs.  (Spinor weight is an additive assignment 
of $+r$ for each $(\tilde\lambda_i)^r$ and $-r$ for each $(\lambda_{i})^r$ 
in a product of terms.) 
The simplest solution to this condition is a quartic polynomial in the 
$\tilde\lambda_i,\lambda_{i}$, linear in $\tilde\lambda_4$ and
$\lambda_1$, with others appearing in the combination 
$\tilde\lambda_i\lambda_{i}$. 
The differential equation then forces a solution of the form 
\begin{equation}
\hat b_3 = 
2P^2  \la 4 | \Slash{5} | 1 \ra   + \alpha (2k_5 \cdot P)    
\la 4 | \Slash{P} | 1 \ra \,.
\end{equation}
The arbitrary coefficient $\alpha$ can easily be fixed to be $-1$
by considering the collinear limit $2-3$. 

Thus we have demonstrated how the action of the holomorphic anomaly on the 
cuts can be used to provide information about $N=1$ supersymmetric
amplitudes.  In general, we obtain differential equations; hence 
fixing the coefficients unambiguously does require the input of 
suitable physical information, such as the collinear limits.  


\section{A term in $A^{N=1\ {\rm chiral}}(1^-,2^-,3^-,4^+,\ldots  ,n^+)$}

As a further example let us consider the $n$-point amplitude 
$A^{N=1\ {\rm chiral}}(1^-,2^-,3^-,4^+,\ldots  ,n^+)$
and deduce some of its integral function coefficients. 
Consider the cut analogous to the previous case $C_{5\cdots n1}$ which is
\begin{equation}
C_{5\cdots n 1} 
={i K\over 2} \int \dlips
        \frac{\spb4.{\ell_2}\spa1.{\ell_1}}{\spb2.{\ell_2}\spa5.{\ell_1}} \,,
\end{equation}
where now
\begin{equation}
K = \frac{\langle 4 | \Slash{P}_{234} | 1 \rangle^2}{\spb2.3
\spb3.4\spa5.6\spa6.7 \cdots \spa{n}.1  s_{234} }
\end{equation}
Notice that on the cut the integrand is
\begin{equation}
\frac{\spb4.{\ell_2}\spa1.{\ell_1}}{\spb2.{\ell_2}\spa5.{\ell_1}}
= { \la 4^+ | \lsl_2 | 2^+ \ra \la 5^+ | \lsl_1 | 1^+ \ra
   \over \la 2^+ | \lsl_2 | 2^+  \ra \la 5^+ | \lsl_1 | 5^+ \ra }
= - { \la 4^+ | \lsl_2 \twosl \Psl_{234} \fivesl \lsl_1 | 1^+ \ra
   \over \la 2^+ | \Psl_{234} | 5^+ \ra  (\ell_2-k_2)^2 (\ell_1+k_5)^2 }
\,.
\label{nCutIntegrand}
\end{equation}
The two propagators in \eqn{nCutIntegrand}, plus the two cut
propagators, make up a cut box integral.  However,
in the numerator of \eqn{nCutIntegrand}, we can anticommute
$\lsl_2$ and $\lsl_1$ toward each other, to get

\begin{eqnarray}
\frac{\spb4.{\ell_2}\spa1.{\ell_1}}{\spb2.{\ell_2}\spa5.{\ell_1}}
&=& { \la 4^+ | \Psl_{234} \fivesl \lsl_1 | 1^+ \ra
   \over \la 2^+ | \Psl_{234} | 5^+ \ra  (\ell_1+k_5)^2 }
 + { \la 4^+ | \twosl \lsl_2 \Psl_{234} | 1^+ \ra
   \over \la 2^+ | \Psl_{234} | 5^+ \ra  (\ell_2-k_2)^2 }
\nonumber \\ && \hskip0.cm
- { \la 4^+ |  \twosl \lsl_2 (\lsl_1 + \lsl_2) \lsl_1 \fivesl  | 1^+ \ra
   \over \la 2^+ | \Psl_{234} | 5^+ \ra  (\ell_2-k_2)^2 (\ell_1+k_5)^2 }
\,,
\label{nCutIntegrand2}
\end{eqnarray}
where we used $P_{234} = \ell_1 + \ell_2$ in the last term, making it
clear that it vanishes.  Thus the cut reduces to a sum of two
cut linear triangles, or in other words,
\begin{equation}
\eqalign{ A^{N=1,{\rm chiral}}& (1^-,2^-,3^-, 4^+,5^+,\ldots,n^+) 
\cr
& = 
-{i \over 2} \biggl[   b_2{ L_0 [ s_{234} /s_{34} ] \over s_{34}} 
+ b_3{ L_0 [ s_{234} /s_{6\ldots1} ]
\over s_{6\ldots1}}\biggr] +
\hbox{\rm terms not contributing to the $C_{5\cdots n 1}$ cut,}
\cr}
\label{nAmpDecomp}
\end{equation}
where $s_{6\cdots1} \equiv (k_6+k_7+\cdots+k_n+k_1)^2$.
Acting upon $C_{5\cdots n 1}$ as before with $\la \bar F_{234} , \bar\eta \ra$,
we obtain
\begin{equation}
\la \bar F_{234} , \bar\eta \ra
C_{5\cdots n 1} 
= - i \pi 
K 
{{P^2 \spb2.4  \spb3.4 \la2,\bar \eta\ra } 
\over
{(2k_2\cdot P)^2}}
{ \la 2 |\Slash P|1 \ra \over \la 2 |\Slash P|5 \ra  } \,.
\end{equation}
Applying exactly the same steps as before we have a trial solution
\begin{equation}
\hat b_2 = - 2 P^2  \la 4 | \Slash 2 | 1 \ra 
+ \alpha (2k_2 \cdot P)    \la 4 | \Slash P | 1 \ra \,,
\end{equation}
where we can fix $\alpha=1$ using collinear limits. 


\section{Conclusions}

We have examined how the holomorphic anomaly acts upon the cuts of
$N=1$ supersymmetric one-loop amplitudes, focusing upon a six-gluon
non-MHV amplitude (calculated by independent methods).  We have shown
that one must take into account the anomaly when acting with the
collinear differential operators on the cuts, in order to match the effect of
the operator acting upon the imaginary part of the amplitude --- as
required by the optical theorem.  As a calculational tool to evaluate
amplitudes, application of the holomorphic anomaly gives differential
equations for the coefficients of the integral functions, unlike the
$N=4$ case where algebraic equations arose.  Since the equations are
differential, their general solution contains homogeneous parts which can be fixed by
the boundary conditions or physical constraints such as collinear limits.

Acknowledgements: We are grateful to Zvi Bern for useful comments
on the manuscript, and to David Kosower for useful conversations. 
L.J.D. and D.C.D. would like to thank the Aspen Center for Physics
for hospitality when this work was initiated.
\vfill\eject


\end{document}